\documentclass[twocolumn,superscriptaddress,showpacs,prl,amsmath,amssymb]{revtex4}

\usepackage{graphicx}
\usepackage{dcolumn}
\usepackage{bm}
\usepackage{hyperref}
\usepackage{color}

\begin{document}

\title{Longitudinal Wobbling Motion in $^{187}$Au}

\author{N. Sensharma}
\author{U. Garg}
\affiliation{Physics Department, University of Notre Dame, Notre Dame, IN 46556, USA}

\author{Q. B. Chen}
\affiliation{Physik-Department, Technische Universit\"{a}t M\"{u}nchen, D-85747 Garching, Germany}

\author{S. Frauendorf}
\author{D. P. Burdette}
\author{J. L. Cozzi}
\author{K. B. Howard}
\affiliation{Physics Department, University of Notre Dame, Notre Dame, IN 46556, USA}

\author{S. Zhu}
\affiliation{National Nuclear Data Center, Brookhaven National Laboratory, Upton, NY 11973, USA}
\author{M. P. Carpenter}
\author{P. Copp}
\author{F. G. Kondev}
\author{T. Lauritsen}
\author{J. Li}
\author{D. Seweryniak}
\author{J. Wu}
\affiliation{Physics Division, Argonne National Laboratory, Argonne, IL 60439, USA}

\author{A. D. Ayangeakaa}
\author{D. J. Hartley}
\affiliation{Department of Physics, United States Naval Academy, Annapolis, MD 21402, USA}

\author{R. V. F. Janssens}
\affiliation{Department of Physics and Astronomy, University of North Carolina Chapel Hill, NC 27599, USA}
\affiliation{Triangle Universities Nuclear Laboratory, Duke University, Durham, NC 27708, USA}

\author{A. M. Forney}
\author{W. B. Walters}
\affiliation{Department of Chemistry and Biochemistry, University of Maryland, College Park, MD 20742, USA}

\author{S. S. Ghugre}
\affiliation{UGC-DAE Consortium for Scientific Research, Kolkata 700 064, India}

\author{R. Palit}
\affiliation{Department of Nuclear and Atomic Physics, Tata Institute of Fundamental Research, Mumbai 400 005, India}

\begin{abstract}
The rare phenomenon of nuclear wobbling motion has been investigated in the nucleus $^{187}$Au. A longitudinal wobbling-bands pair has been identified and clearly distinguished from the associated signature-partner band on the basis of angular distribution measurements. Theoretical calculations in the framework of the Particle Rotor Model (PRM) are found to agree well with the experimental observations. This is the first experimental evidence for longitudinal wobbling bands where the expected signature partner band has also been identified, and establishes this exotic collective mode as a general phenomenon over the nuclear chart.
\end{abstract}

\date{\today}

\pacs{27.70.+q, 23.20.-g, 23.20.En, 23.20.Gq, 21.60.Ev}

\maketitle

The shape of a nucleus, determined via specific characteristic spectroscopic features observed in experiments, is a 
manifestation of the self-organization of a finite fermionic system. Studying the appearance of various shapes  with 
changing of neutron-to-proton ratio or increasing angular momentum reveals new insights into fundamental  principles
governing finite fermionic systems in general. 
The range of shapes that nuclei can assume encompasses spherical symmetry and axial deformation near the ground state,
discussed in the textbooks  \cite{bohr}, or coexistence of both shapes in one nucleus \cite{shape_coex}, in analogy to stereo isomers of molecules.

\begin{figure}[h!]
\centering
\includegraphics[width=0.40\textwidth]{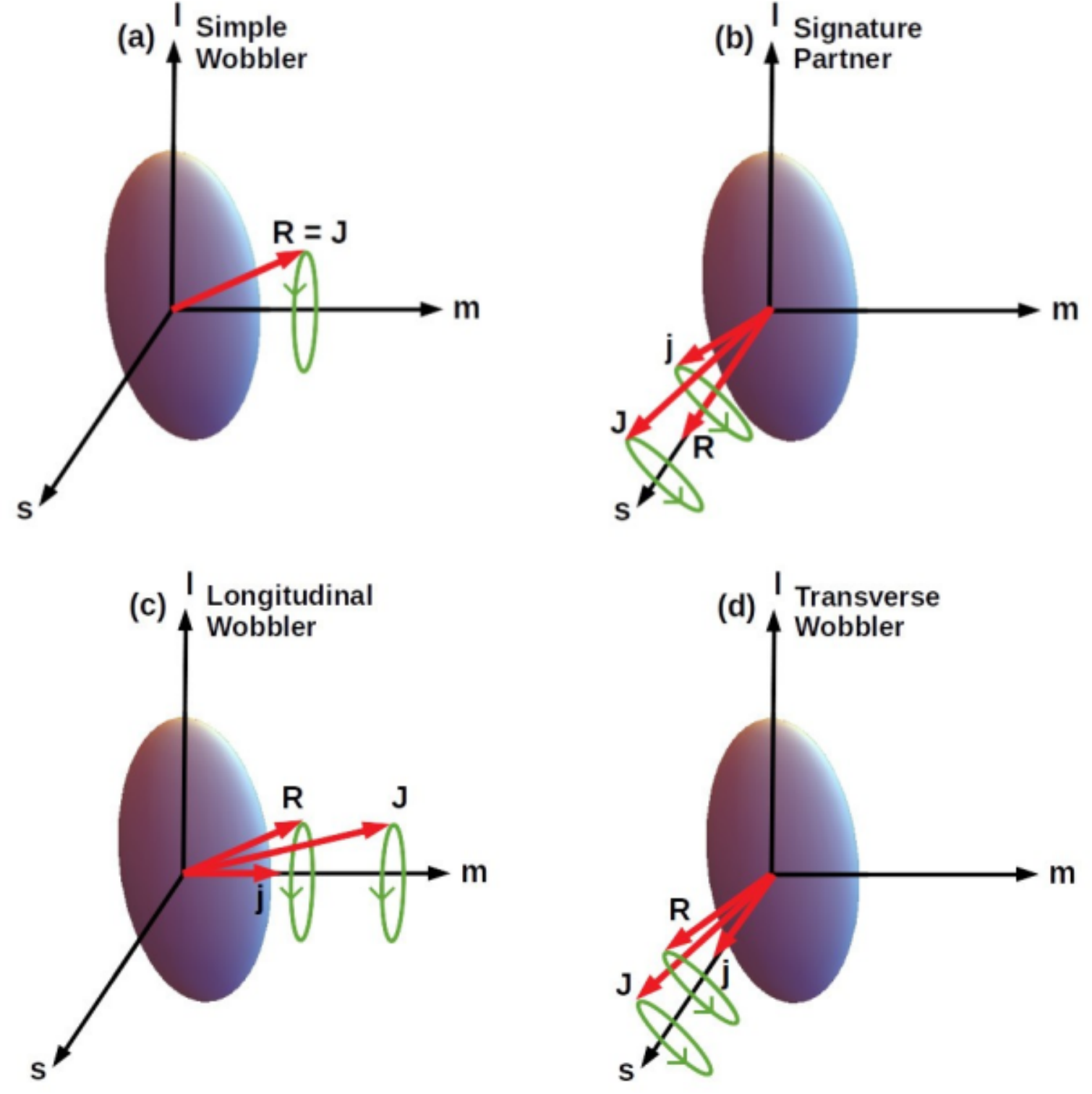}
\caption{\label{f:wobbling}
(Color online) Angular momentum geometry of (a) simple wobbler, (b) signature partner, (c) longitudinal and (d) transverse wobbler
in the body fixed frame, where l, m, and s correspond to the long-, medium-, and short axis, respectively. $R$, $j$ and $J$ are the rotor, odd particle, and total angular momentum, respectively.} 
\end{figure}

Triaxial nuclei (shaped like an ellipsoid with all three axes unequal), being  rare in the ground state  \cite{moeller}, have drawn considerable attention over the years. The experimental observation of this unusual geometry is aided by two fingerprints: chirality and wobbling. Chirality, although not prevalent in nuclear physics because nuclei have rather simple shapes, has been observed, nonetheless, in a number of nuclei across the nuclear periodic table \cite{chirality}. 
Nuclear wobbling motion, the other principal signature of triaxiality, is a collective mode that appears when the moments of inertia ${\cal J}_i$ of all three principal axes of the nuclear density distribution are unequal, which is a clear signal for a triaxial nuclear shape. The mode is well known in classical mechanics. For a given angular momentum, uniform rotation about the axis with the largest  moment of inertia (the m-axis in Fig.~\ref{f:wobbling}) corresponds to minimal energy. At a somewhat larger energy, this axis precesses (wobbles) about the  space-fixed angular-momentum axis $\vec J$. In a quantal system, such as the nucleus (or a molecule), $\vec{J}$ wobbles about the medium (m-) axis in the body fixed frame (as illustrated in Fig. \ref{f:wobbling} (a)). This mode manifests itself in the appearance of rotational bands that  correspond to successive excitations of wobbling phonons, n$_{\omega}$, and alternating signature $\alpha$ = $\alpha_0$ + n$_{\omega}$, which determines the spin sequence I = $\alpha$ + {\em even number}. Adjacent wobbling bands  n$_{\omega+1}$ and n$_{\omega}$ are connected by $\Delta$I  = 1 transitions with a collectively-enhanced  E2 component, which is generated by the wobbling motion of the entire charged body.  

Microscopic calculations give ratios between  the three moments of inertia that are close to the
ratios of irrotational flow \cite{frauendorf2018}. The reason is that collective rotation about a symmetry axis is not possible for a system of 
identical fermions. Accordingly, the medium axis (m-) has the largest moment of inertia, because deviation from axial symmetry is maximal.  Although predicted quite sometime ago \cite{bohr}, there is only fragmentary evidence for simple wobbling (Fig. \ref{f:wobbling} (a)) in even-even nuclei
 \cite{frauendorf2015}.
Instead, wobbling has been demonstrated for a few odd-A nuclei:  $^{105}$Pd \cite{timar}, $^{135}$Pr \cite{matta,two-phonon}, $^{133}$La \cite{sayani}, $^{161}$Lu \cite{bringel}, $^{163}$Lu \cite{odegard, jensen}, $^{165}$Lu \cite{schonwasser}, $^{167}$Lu \cite{amro} and $^{167}$Ta \cite{hartley}. 

All these nuclei have an odd nucleon (neutron in the case of $^{105}$Pd \cite{timar}, and proton for all the other cases) occupying a high-$j$ orbital. Depending on the particle (hole) nature of the odd quasiparticle arising from the bottom (top) of a deformed $j$ shell, its angular momentum gets aligned with the short, s- (long, l-) axes of the triaxial rotor, because this maximizes (minimizes) the overlap 
of its density distribution with the triaxial core, which minimizes the energy. If the quasiparticle arises from the middle of the $j$ shell, it tends to align its angular momentum along the m-axis. As the presence of the odd quasiparticle modifies the wobbling motion considerably, Frauendorf and D\" onau  \cite{transverse} classified it as ``longitudinal wobbling'' (LW) and ``transverse wobbling'' (TW) when, the odd nucleon aligns its angular momentum along the m-axis, or along one of the perpendicular (s- or l-) axes, respectively (Figs. \ref{f:wobbling} (c) and (d)). 
 
The wobbling energy (E$_\textrm{wobb}$), which is the energy of the n$_{\omega}$ = 1 wobbling band relative to the n$_{\omega}$ = 0 band, is defined in Ref. \cite{transverse} as: 
\begin{multline}
E_\textrm{wobb} = E(I, n_{\omega} = 1) - \\ \left[\frac{E(I+1, n_{\omega} = 0) + E(I-1, n_{\omega} = 0)}{2}\right]
\end{multline}

For a qualitative understanding of its $I$-dependence, let us assume that ${\cal J}_\parallel$ refers to the axis of uniform rotation, and the moments of inertia of the two perpendicular axes are equal to ${\cal J}_\bot$ (this is the case for irrotational flow and the triaxiality parameter $\gamma=30^\circ$). In accordance with Frozen Alignment (FA) approximation of Ref. \cite{transverse}, the quasiparticle angular momentum $\vec j$ is assumed to be rigidly aligned with the axis.  With $A_i=1/2{\cal J}_i$, the rotor energy is given by $A_\parallel R^2_\parallel+ A_\bot R_\bot^2$. Exciting the first wobbling quantum corresponds to changing $J_\parallel$ from $I$ to $I-1$, for fixed $I$. The geometry of the precession cones in  Fig. \ref{f:wobbling} implies a change of $R^2_\bot$ from 0 to $\approx 2I$ and of $R^2_\parallel$ from $(I-j)^2$ to $(I-j-1)^2$, which gives an increase in the rotor energy by: 
\begin{equation} 
\label{eq:wobs} 
E_\textrm{wobb} = (A_\bot-A_\parallel)2I + 2\bar{j}A_\parallel,~~~\bar j=j+1/2
\end{equation} 
In case of the simple and the longitudinal wobbler (Fig. \ref{f:wobbling} (a) and (c)) the precession cone revolves about the m-axis with the largest moment of inertia. As  $A_\parallel<A_\bot$, the wobbling energy  $E_\textrm{wobb} $ increases with $I$. For the case of transverse wobbling (Fig. \ref{f:wobbling} (d)), the precession cone revolves about the s- (or l-) axis, which has a smaller moment of inertia than that for rotation about the m-axis. In this case,  $A_\parallel>A_\bot$ and the wobbling energy $E_\textrm{wobb} $ decreases with $I$ until zero, where the mode becomes unstable.

The expression,
\begin{equation}
E_\textrm{wobb}=\sqrt{\left((A_{\bot1}-A_\parallel )2I+2\bar jA_\parallel\right)\left((A_{\bot2}-A_\parallel )2I+2\bar jA_\parallel\right)} 
\end{equation}
obtained in Ref. \cite{transverse} for three different moments of inertia ${\cal J}_{\bot 1}, ~{\cal J}_{\bot 2}, ~{\cal J}_{\parallel}$, is the geometric mean value of the wobbling energies given in Eq. \ref{eq:wobs}.   

It should be understood that the frozen alignment scenario discussed here is an idealization to illustrate the longitudinal and transverse coupling schemes in a transparent way. The odd particle responds to the inertial forces, changing its orientation to a certain degree. Nevertheless, the qualitative classification remains valid. Wobbling is characterized by collectively enhanced $I\rightarrow I-1$, 
E2 transitions from the wobbling to the yrast band, where the wobbling energy increases (decreases) for LW (TW).

The signature-partner bands represent another type of excitation involving a partial de-alignment of the odd particle with respect to its preferred axis (Fig. \ref{f:wobbling} (b)); for those, the connecting $\Delta I$ = 1 transitions are of predominant M1 character, with very little, if any, E2 admixture. 

In all of the cases mentioned above (except $^{133}$La\cite{sayani}), the wobbling bands have been identified as corresponding to TW, because E$_{\textrm{wobb}}$ decreases with increasing angular momentum.
In this Letter, we report on the observation of band structures corresponding to longitudinal wobbling motion in the nucleus $^{187}$Au.
This is the first case of observation of bands corresponding to longitudinal wobbling, clearly distinguished from the associated signature-partner band. Further, these results open up a new mass region, and a different set of orbitals, where this exotic collective motion is established. Occurrence of triaxiality at low spins has been established in this mass region by observation of  chiral band pairs in several nuclei \cite{188Ir, 194Tl, 198Tl} and suggested by large-scale, mean-field calculations (see, for example, Refs. \cite{carpenter,moeller, niksic}). Also, earlier studies of the coupling of an odd number of particles to a rotor had revealed substantial deviations from an axial shape \cite{mtv1975, stefan1977}. 

To populate the levels of interest in $^{187}$Au, a $^{19}$F beam was used with an enriched $^{174}$Yb target (13 mg/cm$^{2}$-thick foil with a 33 mg/cm$^{2}$ $^{208}$Pb backing) at the ATLAS facility of the Argonne National Laboratory. Data were collected with the Gammasphere array in two separate runs using the same beam and target combination. For the first run,  a total of 57 Compton-suppressed Germanium detectors of the Gammasphere array were employed  and the beam energy was 105 MeV. For the second, the number of  detectors was 73, and the beam energy 115 MeV. Data were acquired in the triple-coincidence mode, with the combined total of three- and higher-fold $\gamma$-ray coincidence events being 1.08 $\times$ 10$^{9}$. 

\begin{figure}[h!]
\centering
\includegraphics[width=0.45\textwidth]{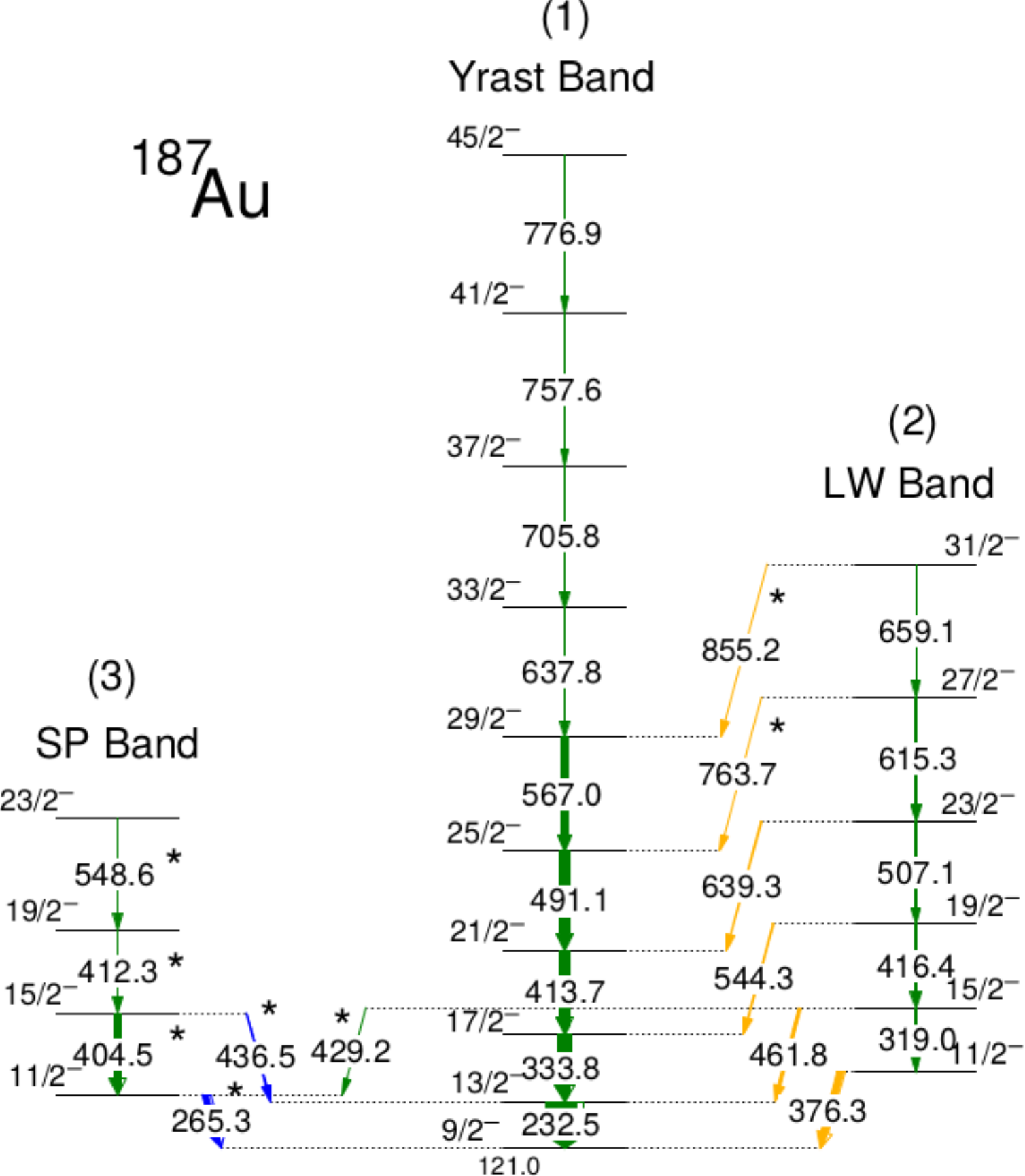}
\caption{\label{f:level_scheme}
(Color online) Partial level scheme of $^{187}$Au relevant to the focus of present work. Shown are the Yrast band, the n$_{\omega}$ = 1 wobbling band (LW) and the signature partner band (SP). Newly identified transitions are marked with an asterisk (*). All the stretched E2 transitions are shown in green, pure M1 transitions in blue, and the M1 + E2 mixed transitions in gold color. The lowest level shown is a 9/2$^{-}$ isomeric level with E$_\text{x}$ = 121.0 keV.} 
\end{figure}

\begin{figure}[h!]
\centering
\includegraphics[width = 0.45\textwidth]{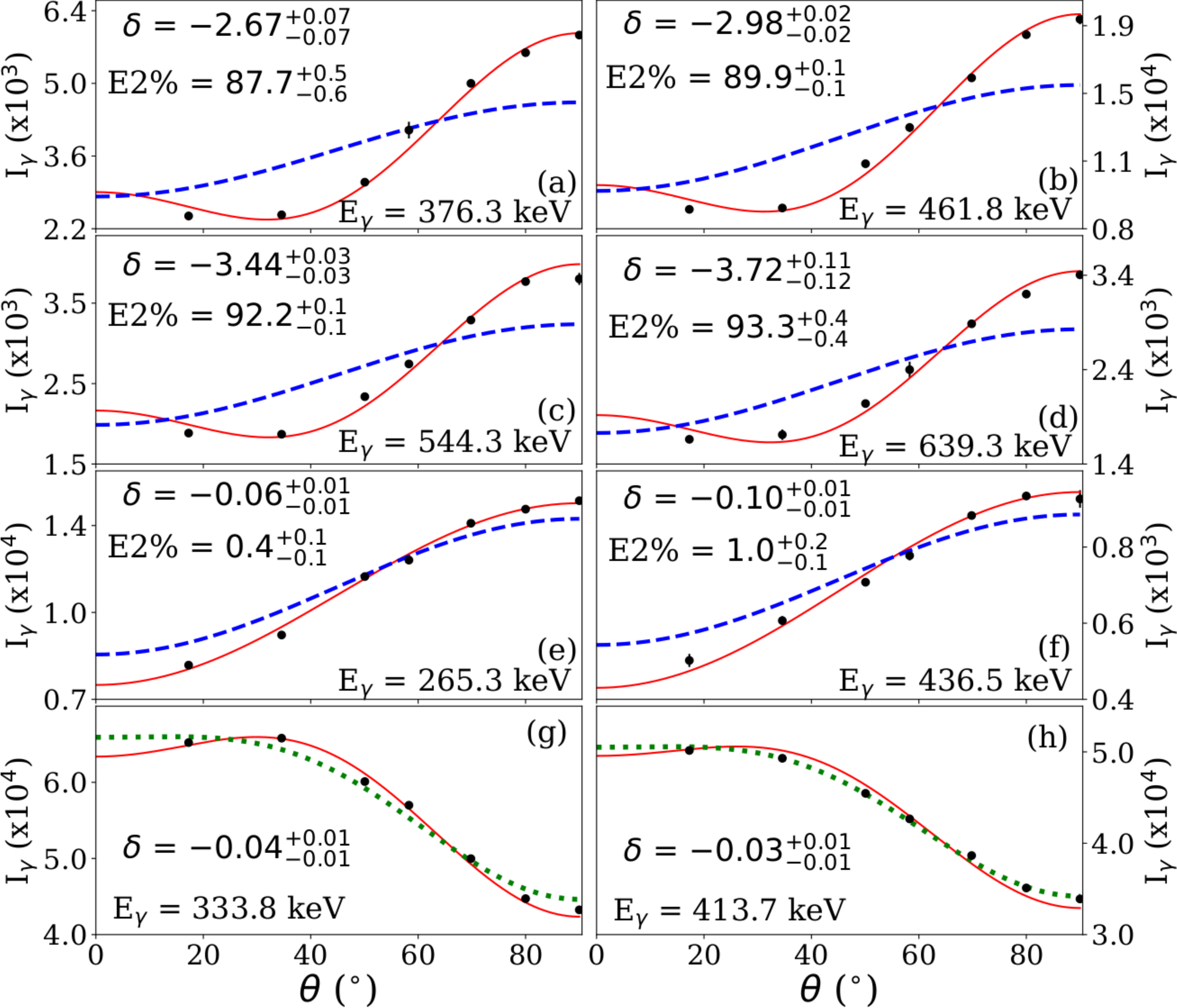}
\caption{\label{f:angdis}
(Color online) Angular distribution plots for the four lowest LW $\to$ Yrast linking transitions [(a)--(d)], two SP $\to$ Yrast linking transitions [(e),(f)], and two Yrast in-band transitions [(g),(h)]. The experimental points are shown by black squares and the solid red lines are fits to the angular distributions. The dashed blue lines [in (a)--(f)] and the dotted green lines [in (g),(h)] represent the expected angular distribution for pure $\Delta$I = 1, M1 and $\Delta$I = 2, E2 transitions respectively.} 
\end{figure}

To take advantage of higher statistics, the data from both measurements were combined, and the analyses performed using the \texttt{RADWARE} suite of codes \cite{radford}. Energy and efficiency calibrations were performed for the added data set and the calibrated data was sorted into $\gamma$-$\gamma$ coincidence matrices and $\gamma$-$\gamma$-$\gamma$ coincidence cubes. A partial level scheme for $^{187}$Au relevant to the focus of this work is presented in Fig. \ref{f:level_scheme}; additional information on the level structure, along with details of the coincidence relationships, as well as the relevant coincidence spectra, will be presented in a forthcoming publication \cite{nirupama3}. The arrangement of Gammasphere detectors into 17 different angular rings around the beam line has enabled high statistics angular distribution measurements for the relevant transitions of Fig. \ref{f:level_scheme}. The analysis procedure followed for these measurements is the same as that described in Refs. \cite{matta,two-phonon}. The validity of the method has been established by examining the angular distributions for two known stretched E2 transitions (333.8- and 413.7-keV) in the Yrast band (shown in Figs. \ref{f:angdis} (g) and (h)). As expected, the mixing ratios extracted for these stretched E2 transitions are extremely small: $\delta$ = -0.04(1) and -0.03(1), respectively. Further details, including a discussion of the various factors that might affect the extraction of final results---efficiency corrections, spin alignment, attenuation coefficients etc.---will be provided in Ref. \cite{nirupama3}. 

Spins and parities of the bandheads as well as some low-lying levels of Bands (1) and (2) had been established previously \cite{deleplanque,bourgeois}. In addition to these, the present work has identified a new band [Band (3)] built on an 11/2$^{-}$ state at 386.3 keV. The spins and parities for the levels in Band (3) have been assigned on the basis of angular distribution measurements as well as coincidence relationships, details of which will be provided in Ref. \cite{nirupama3}. 

Band (2) in Fig. \ref{f:level_scheme} is found to decay to Band (1) (the Yrast band) via six $\Delta I$ = 1 transitions. Previous work \cite{bourgeois2,johansson} has identified this band as the unfavored signature partner of Band (1). However, based on the high-statistics angular distribution measurements for the connecting transitions between the two bands, this sequence has been identified in the present work as the first wobbling (n$_{\omega}$ = 1) band. Figs. \ref{f:angdis} (a) -- (d) provide the angular distributions for the four lowest n$_{\omega}$ = 1 $\to$ Yrast (n$_{\omega}$ = 0) connecting transitions. The mixing ratio, $\delta$, and the percentage of E2 mixing are noted on each plot. These transitions have an 87\%--93\% E2 component, clearly identifying them as $\Delta I$ = 1, E2 in nature, which is the hallmark of wobbling bands \cite{odegard}. The present work has not been able to identify any other wobbling bands corresponding to higher phonon numbers (n$_{\omega}$ = 2 and above). The absence of these bands can be attributed to reasons discussed previously in Ref. \cite{transverse}.

Band (3) is found to decay to the yrast band via two $\Delta I$ = 1 transitions (265.3- and 436.5-keV). The angular distributions for these transitions (Figs. \ref{f:angdis} (e) and (f) )
reveal a very small E2 component ($\approx$ 0.4\% and 1.0\% E2 admixture, respectively), identifying these transitions as being essentially of a pure M1 character. Angular distribution measurements for the in-band transitions have revealed a stretched E2 character. Moreover, a $\Delta$I = 2 crossover transition (429.2-keV) was also identified, connecting the 15/2$^{-}$ level in Band (2) to the 11/2$^{-}$ level in Band (3). The spin and parity of the 15/2$^{-}$ level in Band (2) have been established previously  \cite{deleplanque, bourgeois}. The observed $\Delta I$ = 2 nature of the in-band transitions, as also of the 429.2-keV transition, along with the pure $\Delta I$ = 1 nature of the two connecting transitions (265.3- and 436.5-keV) have led to the spin and parity assignments in Fig. \ref{f:level_scheme}. Band (3), with its two almost pure M1 connecting transitions to the Yrast band, has, thus, been identified as the unfavored signature partner (SP) of the Yrast band.

\begin{figure}[h!]
\centering
\includegraphics[trim=10mm 1mm 1mm 5mm, width = 0.42\textwidth]{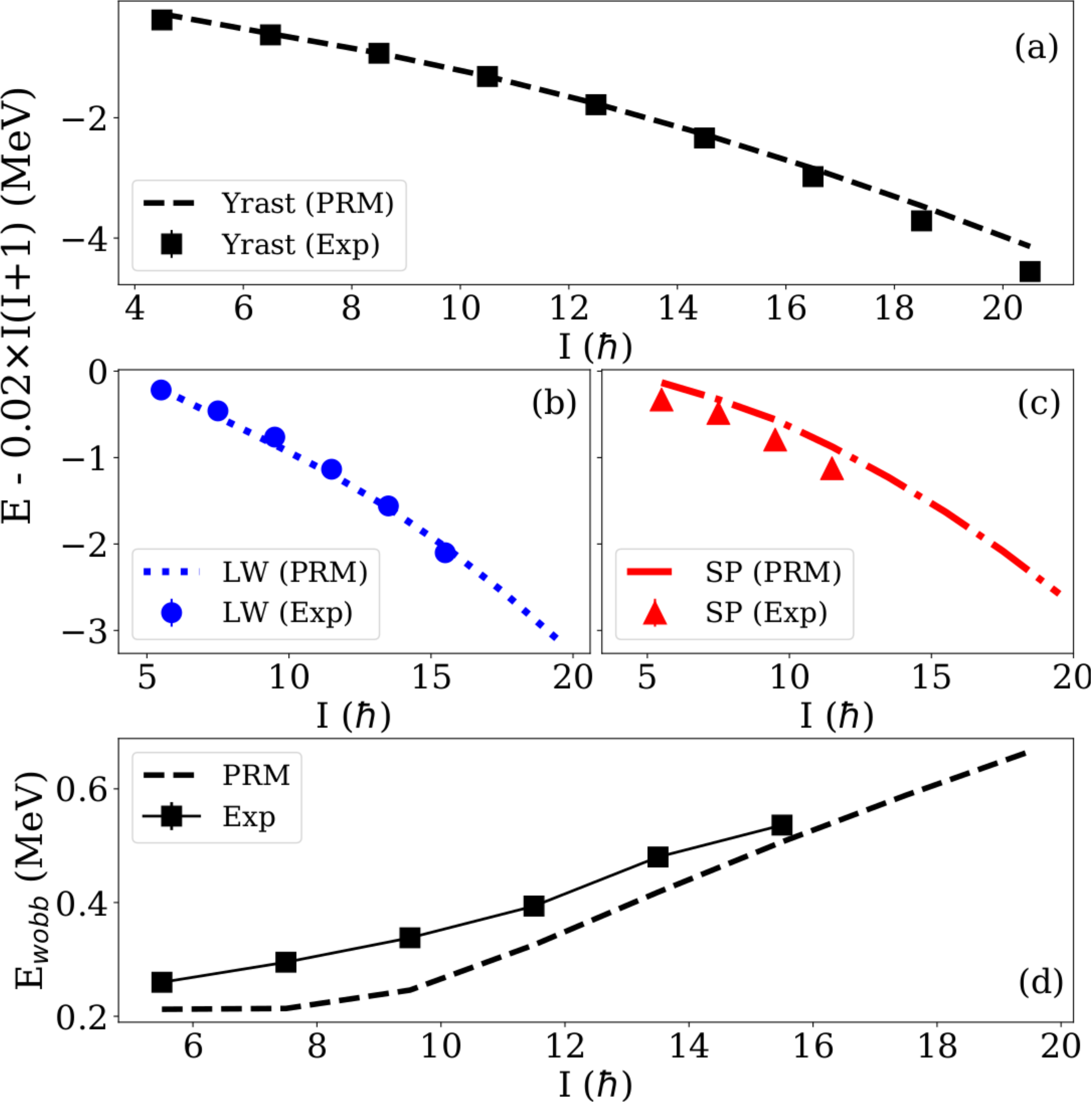}
\caption{\label{f:level_energy}
(Color online) Experimental level energies minus a rotor contribution for the (a) Yrast (b) LW and (c) SP bands. (d) Wobbling energy plot for the LW band as a function of spin. Also shown for comparison are results from PRM calculations (dashed, dotted, and dash-dotted lines). The line through the experimental points in (d) is drawn to guide the eye.} 
\end{figure}

Fig. \ref{f:level_energy} (d) displays the variation of E$_\textrm{wobb}$ with spin. The increasing trend clearly identifies $^{187}$Au as a longitudinal wobbler. This is different from all the other known wobblers, which have been identified as being of the transverse type. Thus, $^{187}$Au represents the first clear observation of longitudinal wobbler motion in nuclei. 

\begin{figure}[h!]
\centering
\includegraphics[width = 0.45\textwidth]{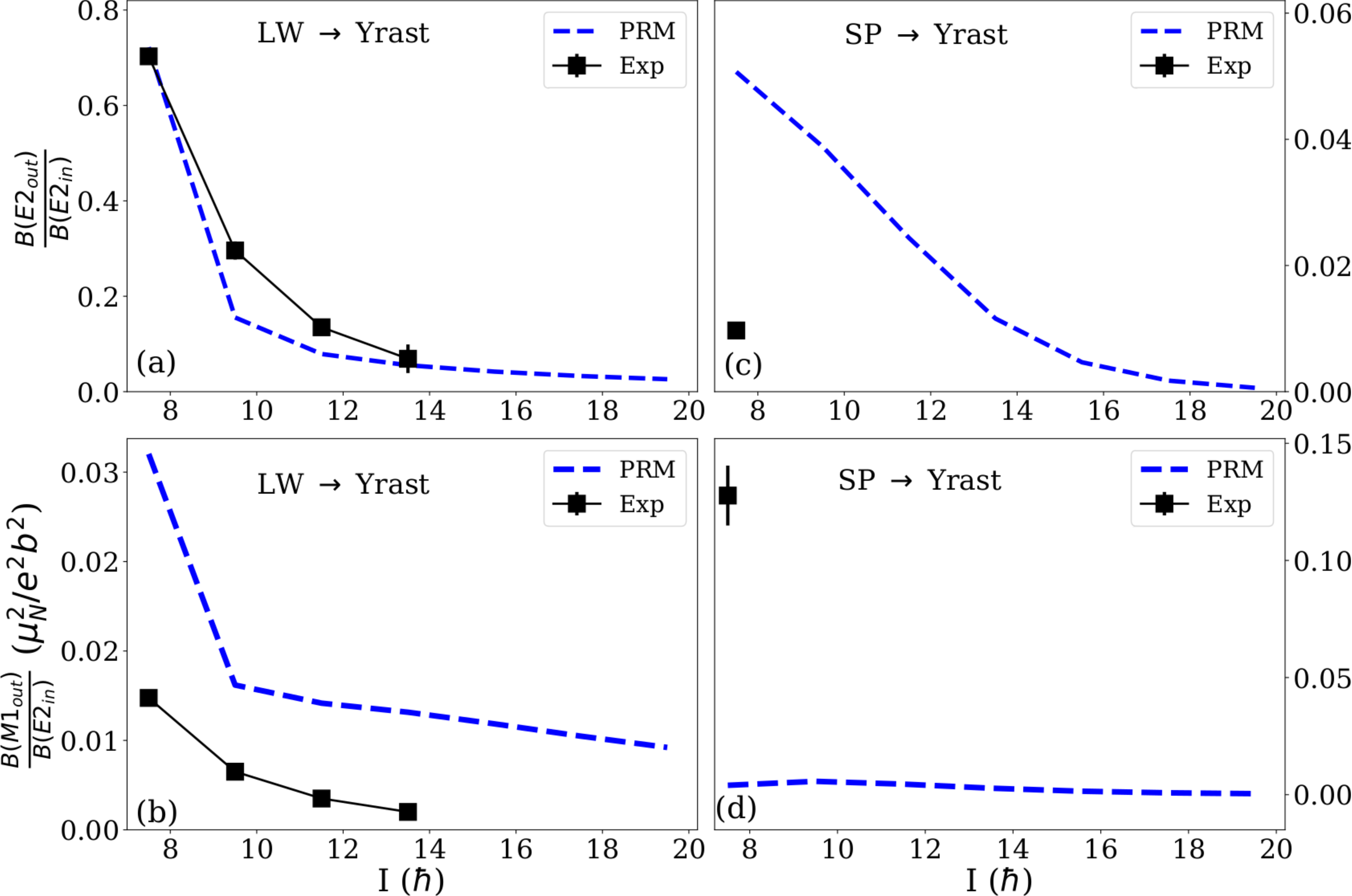}
\caption{\label{f:trans_prob}
(Color online) (a) B(E2)$_\text{out}$/B(E2)$_\text{in}$ ratios \textit{vs.} spin and (b) B(M1)$_\text{out}$/B(E2)$_\text{in}$ \textit{vs.} spin for the LW $\to$ Yrast connecting transitions. (c) and (d): same as (a) and (b), but for SP $\to$ Yrast connecting transitions; the vertical scale for these is on the right side. The experimental points are shown by black squares connected by black solid line and the blue dashed lines correspond to results from PRM calculations.} 
\end{figure}

To further understand the nature of wobbling, we have carried out calculations  in the framework of the Particle Rotor Model (PRM) \cite{transverse, prm1, prm2} for the $h_{9/2}$ band structures, with the deformation parameters $\beta=0.23, \gamma=23^\circ$, the pairing  gap $\Delta= 0.88~\textrm{MeV}$, and the chemical potential  located at $\lambda=-1.32~\textrm{MeV}$,  0.38 MeV below the second level of the $h_{9/2}$ shell. The $h_{9/2}$ proton is described by a single-$j$ shell Hamiltonian. Including the $f_{7/2}$ orbital into the PRM calculations provided results that agree with Figs.  \ref{f:level_energy} and \ref{f:trans_prob}
within the shown accuracy, because the admixture of the $f_{7/2}$ proton to the eigenstates is lower than 5\%.  
We have also carried out cranking calculations based on the configuration-fixed covariant density functional PC-PK1  \cite{pk1,meng,meng2,meng3,meng4}. The equilibrium deformations changed only slightly from $\beta=0.28,~\gamma=22^\circ$ at $I=9/2$ to $\beta=0.28,~\gamma=25^\circ$ at $I=29/2$, which justifies the assumption of a constant deformation for the PRM calculations. The analogue calculations for the $^{186}$Pt core provided quite similar equilibrium deformations. As input for the PRM, we used the smaller deformation $\beta=0.23, \gamma=23^\circ$ found by the HFB-D1S calculations for $^{186}$Pt \cite{HFB-D1S} 
 because it accounts for the experimental values of B(E2,$13/2^-\rightarrow 9/2^-$) = 1.49 $(eb)^2$ for $^{187}$Au \cite{nds_au} and B(E2,$2^+ \rightarrow 0^+$) = 0.84 $(eb)^2$ for $^{186}$Pt \cite{nds_pt}. The moments of inertia are of the irrotational-flow type: $\mathcal{J}_k=\mathcal{J}_0\sin^2(\gamma-2k\pi/3)$, with $\mathcal{J}_0=38.0~\hbar^2/\textrm{MeV}$.
  The PRM calculations locate the chemical potential in the middle of the $h_{9/2}$ shell. Frauendorf and D\" onau \cite{transverse} had predicted the appearance of LW for quasiparticles from half-filled shells. 

Figures \ref{f:level_energy} (a) -- (c) display the experimental level energies minus a rotor contribution for the Yrast, LW and the SP bands. The general agreement between the experimental level energies and those calculated in the PRM is satisfactory. The wobbling energy, as predicted by PRM (shown in Fig. \ref{f:level_energy} (d)), reproduces the increasing trend, but the value is slightly underestimated as compared to the experiment. 

Also indicative of wobbling bands is a high reduced E2 transition probability, B(E2), for the n$_{\omega}$ = 1 $\to$ n$_{\omega}$ = 0 connecting transitions. In Figs. \ref{f:trans_prob} (a) and (b), we present the ratios of the transition probabilities B(E2)$_\text{out}$/B(E2)$_\text{in}$ and B(M1)$_\text{out}$/B(E2)$_\text{in}$, respectively, for these connecting transitions. The measured B(E2)$_\text{out}$/B(E2)$_\text{in}$ ratios are large 
(up to $\sim$ 0.7), indicating that the band exhibits the character of a collective quadrupole excitation, which further strengthens our argument that Band (2) is, indeed, the wobbling band. PRM calculations reproduce the B(E2)$_\text{out}$/B(E2)$_\text{in}$ ratios well but the B(M1)$_\text{out}$/B(E2)$_\text{in}$ ratios are overestimated somewhat (note the very different vertical scales in (a) and (b)).

The results for the transitions from the SP to the Yrast band are presented in Figs. \ref{f:trans_prob} (c) and (d). The B(E2)$_\text{out}$/B(E2)$_\text{in}$ ratios for these transitions are much smaller than those for LW $\to$ Yrast linking transitions, which further supports the wobbling and signature-partner interpretations for the LW and SP bands. The PRM calculations overestimate the B(M1)/B(E2)$_\text{in}$ ratios
for the LW band and underestimate it for the SP bands. This may be  attributed to an incorrect  mixing between 
the wobbling and SP states, which is  sensitive  to the excitation energies and the ratios between the moments of inertia. 

In summary, we have observed wobbling motion in the nucleus $^{187}$Au. Two rotational bands have been identified as corresponding to n$_{\omega}$ = 0 and n$_{\omega}$ = 1  wobbling-bands pair. 
The signature partner of the n$_{\omega}$ = 0 (yrast) band has also been identified. An increasing wobbling energy, E$_\textrm{wobb}$, with spin establishes $^{187}$Au as the first nucleus in which longitudinal wobbling motion has been observed and clearly distinguished from the signature-partner band. Results from PRM calculations are in good agreement with experimental observations. These results open the A$\sim$190 region as a new arena where this exotic collective mode has now been observed and establish this mode as a general phenomenon over the nuclear chart encompassing many different nuclear orbitals. Continuing experimental efforts are warranted to explore this behavior further. 

UG acknowledges travel support from CUSTIPEN (China-U.S. Theory Institute for Physics with Exotic Nuclei) which was instrumental in the initiation of theoretical collaboration with QBC. This work has been supported in part by the U.S. National Science Foundation [Grants No. PHY-1713857 (UND) and No. PHY-1203100 (USNA)], and by the U. S. Department of Energy, Office of Science, Office of Nuclear Physics [Contract No. DE-AC02-06CH11357 (ANL), No. DE-FG02-95ER40934 (UND), No. DE-FG02-97ER41033 (UNC), DE-FG02-97ER41041 (TUNL), No. DE-FG02-94ER40834 (Maryland), and No. DE-SC0009971(CUSTIPEN)]. The work of QBC was supported by Deutsche Forschungsgemeinschaft (DFG) and National Natural Science Foundation of China (NSFC) through funds provided to the Sino-German CRC 110 ``Symmetries and the Emergence of Structure in QCD" (DFG Grant No. TRR110 and NSFC Grant No. 11621131001). This research used resources of ANL's ATLAS facility, which is a DOE Office of Science User Facility.

\bibliography{UND_187Au_rev2}

\end{document}